\documentstyle[amssymb,epsf,pra,aps]{revtex}

\begin{document}

\title{On the Equivalence Principle in Quantum Theory \footnote{Will appear in {\it General Relativity and Gravitation (1996)}}}

\author{Claus L\"ammerzahl\cite{AAAuth}}

\address{Laboratoire de Gravitation et Cosmologie Relativiste, Universit\'e Pierre et Marie Curie, \\ CNRS/URA 679, F - 75252 Paris Cedex 05, France \\
and \\
Fakult\"at f\"ur Physik der Universit\"at Konstanz, Pf 5560 M 674, \\ D - 78434 Konstanz, Germany}

\maketitle

\begin{abstract}
The role of the equivalence principle in the context of non-relativistic
quantum mechanics and matter wave interferometry, especially atom beam
interferometry, will be  discussed. A generalised form of the weak
equivalence principle which is capable of covering quantum phenomena
too, will be proposed. It is shown that this generalised equivalence
principle is valid for matter wave interferometry and for the dynamics
of expectation values. In addition, the use of this equivalence
principle makes it possible to determine the structure of the
interaction of quantum systems withgravitational and inertial fields.
It is also shown that the path of the mean value of the position
operator in the case of gravitational interaction does fulfill this
generalised equivalence principle.
\end{abstract}

\pacs{....}

\section{Introduction}

While the equivalence principle (EP) is a well defined principle in classical physics (see e.g. Will \cite{Will93}) it is not clear whether there is an analogous principle in the quantum domain. 
The main problem seems to be that for the formulation of EPs in the classical domain one has to use local notions like path (the path of a structureless point particle does not depend on the mass of the particle) or point (gravity can be transformed away in one point) while quantum physics is genuinely non-local (fields in general are spread over all space, expectation values are integrals over a whole $t = const.$ hypersurface, etc.). 

Beside this general issue there are many arguments which indicate that there is no EP in quantum theory:

\begin{enumerate}
 
\item Because the phase shift of a neutron interference experiment in a gravitational field $\delta\phi = m g A/\hbar v$ depends on the mass \cite{COW75}, the weak EP is regarded as not being valid for quantum phenomena \cite{Greenberger68,GreenbergerOverhauser79}. 

\item On the other hand, the COW experiment \cite{COW75} together with the experiment of Bonse and Wroblewski \cite{BonseWrob83} may be used to verify the strong EP in the sense that gravitational acceleration is equivalent to the acceleration of a reference system (for an extensive discussion see \cite{HehlLemkeMielke91,AHL92,LemkeMielkeHehl94}). 
However, this statement is exactly true only for vanishing curvature, i.e.\ for uniform acceleration or homogeneous gravitational fields. 
The influence of a non-linear Newtonian potential (or of curved space-time) on the interference pattern cannot be transformed away using accelerated frames, (see below). 

\item The wave function solving the Schr\"odinger equation in a homogeneous gravitational field or in an accelerated frame depends on the mass \cite{GreenbergerOverhauser79}. 
This feature is to be expected because the equation of motion, contrary to the equation of motion for point particles, depends on the mass. 
This is true in the relativistic as well as in the non-relativistic domain. 

\item The EP is certainly not fulfilled for particles with internal structure like spin. 
Using a WKB-approximation of the Dirac equation in Riemann-Cartan space-time it has been shown \cite{Audretsch83} that the spin couples to the space-time curvature and the torsion so that the particle's path and energy depend on the spin direction. 

\item In addition, within another context like a causal interpretation of quantum  mechanics, it has also been shown that quantum theory does not obey the EP \cite{Holland89}. 

\end{enumerate}

Contrary to all statements that there is no EP in the quantum domain, it has been remarked by Treder \cite{Treder82} that a paradoxical situation will arise if on the one hand all equations of motion (field equations for quantum matter, e.g. Klein-Gordon, Dirac equation) are covariantly coupled to gravity (minimal coupling procedure) and on the other hand the influence of gravity on quantum matter may be mass dependent. 
Since the Schr\"odinger equation is the non-relativistic limit of the Klein-Gordon- or the Dirac equation, this statement holds for this equation too. 
Our considerations will support the point of view that the effects of gravity on quantum matter are essentially mass independent if written in an appropriate way.   

Of course, characteristic quantum properties like the Compton wavelength or the width of the interference fringes of matter waves which pass through a double slit do depend on the mass of the matter wave. 
However, this is not the problem we are dealing with. 
Our question is whether there are {\it quantum effects which are universally influenced by the interaction with gravitation}. 
Thereby we refer to quantum effects as the interference pattern in matter wave interferometry or the path of the mean value of the position operator. 
By universal influence we mean an influence which is independent of characteristic properties of the quantum matter like mass. 
It will turn out in the following that contrary to the remark of Treder effects like the phase shifts in atomic beam and neutron interferometry in general do depend on the mass of the particles, so that the EP is not fulfilled. 
The reason is that especially in quantum mechanics it is necessary to disturb the quantum system by some input in order to measure its properties. 
That means that in matter wave interferometry we have to take into account the quantum nature of the interaction of the beam splitters and mirrors with the matter field. 
However, it proves to be possible and useful to introduce the notion of an {\it input independent result}. 
This notion extracts from the measured effect those parts which are characteristic for the `pure' interaction of the quantum matter with external fields. 
It gives a mass independend result in the case of a gravitational interaction. 
This kind of result is independent of the input of the measuring apparatus used into the quantum system.  

In the following we treat spinless matter fields obeying the  Schr\"odinger equation. 
We will not treat relativistic effects. 
Thereby we restrict our considerations to first quantisation. (We are not dealing with the EP in quantum field theory, see e.g. \cite{GinzburgFrolov87} where it is shown that a homogeneous gravitational field and a uniform acceleration are equivalent if one takes correct boundary conditions.) 

Our proposition is that quantum mechanics fulfills a slightly modified EP which we call a Quantum Equivalence Principle (QEP). 
It states that for each given initial state the input independent result of a physical experiment is independent of the mass or charge of the quantum system. 
Thereby we lay stress on an {\it operational realisation} of the QEP in the quantum domain. 
In addition, the QEP is valid in arbitrarily curved space-time. 
The QEP is a statement about experimental results and not about the structure of a field equation. 
The latter can be derived from the experimental results and the QEP. 
In the case of atom beam and neutron interferometry we discuss and analyse the extent to which the EP and the QEP is valid. 
We show that the mass dependence of quantum phenomena like the modification of the interference fringes caused by the gravitational field is due to the quantum nature of the interaction of the quantum systems with the `quantum beam splitters' and `quantum mirrors' which are given by laser waves and crystals. 
For quantum phenomena it is not possible to formulate a strong EP. 

The validity of the QEP implies that the influence of gravitation on quantum matter is of geometrical nature. 
This implies that one can define fields which couple {\it universally} to quantum systems, that is, independently of its properties. 
Using the QEP the other way round, the validity of the QEP implies that it is possible to {\it reconstruct} the strucure of the coupling of gravity to matter field equations. 
This means that the EP or QEP acts as a selection rule for geometrical interactions which are called gravitation. 
This is totally analogous to the approach in the case of point particles where the EP is used to derive the coupling of gravity to structureless point particles which result in the geodesic equation in the case of relativistic mechanics or the equation of motion $\mbox{\boldmath$a$} = - \nabla U$ in Newtonian mechanics. 

\medskip
In the following we will briefly discuss the formulation and implications of the EPs in classical physics. 
Then we introduce the notion of input independent results of experiments which is used to formulate a QEP. 
Thereby we lay emphasis on atom beam interferometry which was first proposed theoretically by Bord\'e \cite{Borde89} for a spatial interaction geometry and subsequently realised \cite{Riehleetal91}. 
Here we theoretically treat the atom beam interferometer used by Kasevich and Chu \cite{KasevichChu91,KasevichChu92} based on a temporal interaction  geometry. 
(Due to the great accuracy which may be achieved by atom beam interferometry, this device attracted great interest as a possible tool for precise experiments in the gravitational field to measure the space--time curvature \cite{AudretschMarzlin93} or to test gravitational theories \cite{ABL93,La94}.) 
In addition we also treat neutron interferometric experiments using a single crystal of the Bonse-Hart type. 
Compared to neutron interferometers atom beam interferometers are more precise and more flexible in their experimental variability. 

Our QEP will be discussed mainly in the context of matter wave interferometry. 
We show that matter wave interferometric measurements are in accordance with our QEP. 
In addition, we show that the path of the mean value of the position operator also fulfills our QEP. 
This path especially does not depend on the mass of the matter field, so that the path of the mean value of the position operator of atoms, neutrons, electrons etc.\ coincide when subject to gravitational interaction only. 

\section{The EP in classical physics} 

In order to touch on the notions involved, we briefly discuss some features  and implications of EPs in classical physics. 
Classical physics thereby means that we are dealing with structureless point particles. 
The EPs are therefore necessarily {\it local} statements dealing with local properties of particle paths.

The {\it weak EP} consists of two parts: First the universality of free fall which means that the path of the structureless point particles does not depend on their mass or composition. 
This is not enough to describe the path by some ordinary differential equation because the path in the future may depend on its history. 
Because this has never been observed, the weak EP states secondly that, if the position $x^a$ and the initial velocity $\dot x^a$ is given then the path is uniquely determined, $a = 1,2,3$. 
By means of this second statement it is now possible to describe the path by an ordinary differential equation of second order: $\ddot x^a = h^a(x, \dot x)$. 
The trajectory is uniquely determined, and does not depend on the mass or on the composition of the structureless particle. 
The main conclusion is that any effect of gravity like the geodesic deviation is universal and that gravity therefore can be geometrised. 
In the framework of Newtonian mechanics the weak EP amounts to the statement that inertial mass and gravitational mass are equal. 
For the present status of experimental verification of the universality of free fall see \cite{Adelbergeretal94}. 

The {\it strong EP} states that locally all (non-gravitational) physical laws acquire their special relativistic form. 
That means the effect of gravity can be locally transformed away by choosing a suitable reference frame, so that gravity is locally indistinguishable from some non-inertial motion. 

For the path of point particles obeying the strong EP this means that locally there is a coordinate system and a parametrisation so that $d^2 x/d\tau^2 = 0$. For arbitrary coordinates and parameters we now have an equation of the structure $\ddot x^\mu + \Gamma^\mu_{\nu\sigma} \dot x^\nu \dot x^\sigma = \alpha \dot x^\mu$ for some function $\alpha$ depending on the parametrisation of the path. 
The strong EP also applies pointwise for fields. 
Mathematically it is the minimal coupling procedure (see \cite{MTW73,AHL92}).

The strong EP implies the weak EP. 
While the weak EP is necessary and sufficient for the geometrisation procedure of the gravitational interaction, the strong EP fixes the structure of this interaction and is therefore a stronger principle. 
The strong EP is also crucial for treating General Relativity as gauge theory, see \cite{AHL92}. Both EPs use local notions. 
The weak EP is a statement of universality: the effect of gravity is the same for all particles, it does not depend on the properties of the particles. 
The weak EP does not require that gravity can be transformed away locally. 

Quantum mechanics is a non-local description of matter.  
For non-local phenomena the effects of gravity cannot be transformed away. 
Consequently, to formulate an EP for the quantum domain we have to avoid any reference to locality and to procedures that transform away effects. 
While the strong EP cannot be freed from notions of locality the weak EP may be generalised by stating that the effect of gravity in a certain experimental setup is the same for a whole class of physical systems prepared with the same initial conditions. 
In the following the class of physical systems consists in the quantum systems fulfilling the Schr\"odinger equation.

\section{Formulation of a weak equivalence principle in the quantum domain}

At the beginning we introduce the following notions which are characteristic for physical experiments: 
A point particle or a quantum system which is subject to the experiment is called a {\it physical system}. 
The assignement of a position and velocity at a certain moment for a point particle or a certain initial state for a quantum system is the {\it preparation} of the physical system. 
After preparation the physical system evolves in some {\it physical situation}, that is, in some external fields (Maxwell field, gravity), perhaps with boundary conditions. 
In order to be able to measure properties of the physical systems one has to manipulate the physical system while preparing it or during its evolution by some experimental {\it input} (e.g. throwing light on the point particle to track its path, shooting photons with a well defined wave vector on atoms to transmit momentum etc.). 
To be more clear: by input we mean the strength of the quantity manipulating the physical system, e.g. the value of the wave vector of the photon interacting with the atom. 
In the case of quantum systems this input in general disturbs the evolution of the quantum system and changes its properties. 
The altered properties also give rise to an altered behaviour of the system in the external fields. 
Therefore it is of importance especially for the quantum domain to be able to reformulate the experimental results in such a way that one can extract results  which are not sensitive to the changes of the physical system due to the experimental input. 
The notion which fulfills this demand and which we are going to define below  is the notion of an input independent result. 
This gives information about the interaction of the physical system with the external field which are `pure' and characteristic for the quantum system interacting with an external field or force. 

Let $S$ be a physical system, $I$ the experimental input to this system and $E(S, I)$ a measured effect which depends on the input acting on the physical system. 
For one physical system there may be many types of inputs, $I_1, I_2, \ldots$,  corresponding to the properties of the physical system which is going to be measured. 
The following analysis holds for each type of input. 

We assume that, if there is no input $I = 0$, then there is no effect: $\displaystyle \lim_{I\rightarrow 0} E(S, I) = 0$. 
(If one is given a formula for an effect $E^\prime(S, I)$ which does not fulfill this relation, i.e. $\displaystyle \lim_{I\rightarrow 0} E^\prime(S, I) \neq 0$, then one can define an effect by $\displaystyle E(S, I) := E^\prime(S, I) - \lim_{I\rightarrow 0} E^\prime(S, I)$.) 
For simplicity we also assume that the dependence of $E(S, I)$ on $I$ is $C^\infty$. 
This implies that there is a real number $p$ and another function $\widehat E(S, I)$ with the property $\displaystyle \lim_{I \rightarrow 0}\widehat E(S, I) \neq 0$ and $<\infty$ so that $E(S, I) = I^p \widehat E(S, 0)$. 
We call the limit $\displaystyle E(S) := \lim_{I\rightarrow 0} {{E(S, I)}\over{I^p}}$ the {\it input independent result}. 
It does not depend on $I$ anymore. 
The input independent result describes the effect for vanishing input normalised by the respective input. 

An example for the above procedure from classical physics is the geodesic deviation equation $b^\mu = R^\mu_{\phantom{\mu}\nu\rho\sigma} u^\nu \delta r^\rho u^\sigma$ by means of which the curvature tensor can be determined \cite{AL83}. 
$b^\mu$ is the measured acceleration, $u^\mu$ is the four velocity, and $\delta r^\mu$ is the space-like distance vector between two neighbouring geodesics. 
$\delta r^\rho$ is part of the preparation procedure and represents the `input'. 
The input independent result is the curvature which is determined by $\displaystyle \lim_{r\rightarrow 0} {1\over r} b^\mu = R^\mu_{\phantom{\mu}\nu\rho\sigma} u^\nu \bar r^\rho u^\sigma$ for all directions $\bar r^\rho := \delta r^\rho/r$ with $r := (- g_{\mu\nu} \delta r^\mu \delta r^\nu)^{1/2}$. -- We will use the concept of an input independent result in the context of quantum interference which is described by a closed path in some parameter space. 
Then this concept describes the infinitesimalisation of the closed path. 

Using the notions introduced above we now propose a version of an EP for the quantum domain: 

\bigskip
\noindent {\bf Quantum Equivalence Principle (QEP):} {\it For all given initial states the input independent result of a physical experiment is independent of the characteristic parameters (like mass, charge) of the quantum system}. 

\bigskip
We will make a few general comments on this QEP:

\begin{enumerate}

\item There are two major points used in the formulation of this QEP: the initial state and the input independent result. 
The first one is necessary to get rid of the nonlocality in quantum theory: the initial value of an expectation value is not enough to determine the time evolution of this expectation value. 
Therefore one has to use the initial state which in general is described by a spread-out field. 
The second point is necessary to avoid the quantum nature of the interaction: if one weakens more and more the input acting on the spread-out quantum system, then one gets after normalisation a result which does not depend on the input.  

\item The above formulation of a QEP uses no local terms. 
It is again a universality statement in the sense that an experimental result is the same for all physical systems. 
The universality makes it possible to geometrise the gravitational interaction in the quantum domain. 
We do not require that gravity can be transformed away. 
This is possible only locally and is therefore not a notion which is appropriate for spread-out fields. 
(Of course, a homogeneous gravitational field can be transformed away so that in this case the strong EP is fulfilled. 
However, in this case the curvature vanishes so that we are restricted to a flat space-time.) 

\item The QEP is stated without reference to gravitation. 
This has to be understood in an {\it operational} manner: each interaction which is universal in the above sense can be geometrised and will be called gravitation. 
In this way each equivalence principle (for classical point particles as well as for quantum systems) can single out and define the gravitational interaction.  

\end{enumerate}

We will show in the following that the gravity induced phase shifts of atom beam and neutron interference experiments and the time evolution of expectation values and uncertainties support the above version of a QEP.

\section{The QEP in matter wave interferometry}

\subsection{The QEP in atom beam interferometry}

\subsubsection{The general dynamical equations}

We start with a treatment of three-level atoms used by the Raman light-pulse interferometer \cite{KasevichChu91}. 
The three levels are denoted by $\alpha = 1, 2, i$, where $i$ stands for `intermediate state'. 
We base our considerations on the Hamiltonian ($\widehat{\mbox{\boldmath$p$}}$ is the momentum operator, $\widehat{\mbox{\boldmath$x$}}$ the position operator, bold-face symbols are 3-vectors, indices run from 1 to 3; although not necessary, we distinguish between covariant and contravariant tensor indices and use Einstein's summation convention) 
\begin{equation}
\widehat H = {{{\widehat{\mbox{\boldmath$p$}}}^2}\over{2m}} + \sum_\alpha \omega_\alpha^A \mid \alpha\rangle\langle \alpha\mid - \mbox{\boldmath$d$} \cdot \mbox{\boldmath$E$} + m U(\widehat{\mbox{\boldmath$x$}})\, . \label{Schroedinger}
\end{equation}
$\omega_\alpha^A$ is the energy of the $\alpha$th level of the atom. 
$- \mbox{\boldmath$d$} \cdot \mbox{\boldmath$E$}$ is the usual dipole interaction ($\mbox{\boldmath$d$}$ is the electric dipole operator and $\mbox{\boldmath$E$}$ the electric field) and $m U(\mbox{\boldmath$x$})$ describes the interaction with the gravitational field where $U(\mbox{\boldmath$x$})$ is the Newtonian potential.\footnote{In a later discussion we will distinguish between the inertial mass $m_i$ and the gravitational mass $m_g$ appearing in the kinetic and gravitational interaction part, respectively. 
The fact that in (\ref{Schroedinger}) we take these two masses as equal does not anticipate that the EP in terms of {\it measured} physical effects is fulfilled. 
A possible violation may be due to the nonlocality of the quantum state and the quantum interaction with beam splitters. 
In fact, the result (\ref{MZPhase1}) shows that the EP is not fulfilled (the QEP is).} 
In the following we expand the Newtonian potential to second order $U(\mbox{\boldmath$x$}) = U({\mbox{\boldmath$x$}}_0) - \delta x^a g_a  + {1\over 2} U_{ab} \delta x^a \delta x^b + {\cal O}((\delta x)^3)$ with the acceleration $g_a := - \partial_a U({\mbox{\boldmath$x$}}_0)$ and the second derivative $U_{ab} = \partial_a \partial_b U({\mbox{\boldmath$x$}}_0)$ which can be identified with the components $R_{a0b0}$ of the Riemannian space-time curvature. 
We neglect spontaneous emission and dephasing. 

In atomic beam interferometry \cite{KasevichChu92} we start with a three-level system which effectively will reduce to a two-level system. 
The first step consists in a decomposition of the state vector into plane waves whereby we split the free evolution:
\begin{equation}
\mid\varphi\rangle = \sum_\alpha \int a_{\alpha, {\mbox{\boldmath$\scriptstyle p$}}}(t) e^{- i \left(\omega_\alpha^A + {{p^2}\over{2 m \hbar}}\right) (t - t_0)} \mid\alpha, \mbox{\boldmath$p$}\rangle d^3p
\end{equation}
with $\mid\alpha, \mbox{\boldmath$p$}\rangle = \mid\alpha\rangle\otimes\mid \mbox{\boldmath$p$} \rangle$ where $\widehat{\mbox{\boldmath$p$}} \mid \mbox{\boldmath$p$} \rangle = \mbox{\boldmath$p$} \mid \mbox{\boldmath$p$} \rangle$ and $\mbox{\boldmath$p$}$ is the momentum of the plane wave. 
The quantum system is prepared at $t_0$. 
$\sum_\alpha a_{\alpha, {\mbox{\boldmath$\scriptstyle p$}}}(t_0)$ is the Fourier transform of $\varphi(\mbox{\boldmath$x$}, t_0)$. 
The electric field consists of two counterpropagating waves ${\mbox{\boldmath$E$}} = {\mbox{\boldmath$E$}}_1 \cos({\mbox{\boldmath$k$}}_1 \cdot {\mbox{\boldmath$x$}} - \omega_1 t + \phi_1) + {\mbox{\boldmath$E$}}_2 \cos({\mbox{\boldmath$k$}}_2\cdot {\mbox{\boldmath$x$}} - \omega_2 t + \phi_2)$ with ${\mbox{\boldmath$k$}}_1 \approx - {\mbox{\boldmath$k$}}_2$. 
We assume that ${\mbox{\boldmath$d$}}\cdot{\mbox{\boldmath$E$}}_{1/2}$ only induces transitions between $\mid 1\rangle$ and $\mid i\rangle$, and $\mid 2\rangle$ and $\mid i \rangle$, and that the rotating wave approximation is valid. 

The interference experiment starts with the preparation of the state $\mid 1, \mbox{\boldmath$p$}\rangle$. 
Projecting successively the above Schr\"odinger equation on this state, on $\mid i, \mbox{\boldmath$p$} + \hbar {\mbox{\boldmath$k$}}_1\rangle$ and on $\mid 2, \mbox{\boldmath$p$} + \hbar ({\mbox{\boldmath$k$}}_1 - {\mbox{\boldmath$k$}}_2)\rangle$ gives the time evolution of the respective coefficients: 
\begin{eqnarray}
\dot a_{1, {\mbox{\boldmath$\scriptstyle p$}}} & = & {i\over 2}\Omega_{1i}^* e^{i \Delta_1 (t - t_0)} a_{i, {\mbox{\boldmath$\scriptstyle p$}} + \hbar {\mbox{\boldmath$\scriptstyle k$}}_1} \nonumber\\
& & - {i\over\hbar} e^{{i\over\hbar} {{p^2}\over{2m}} (t - t_0)} m \left(U - i \hbar g_a {\partial\over{\partial p_a}} - \hbar^2 U_{ab} {\partial\over{\partial p_a \partial p_b}} \right) \left(a_{1, {\mbox{\boldmath$\scriptstyle p$}}} e^{ - {i\over\hbar} {{p^2}\over{2m}} (t - t_0)}\right)\label{Dyna1}\\
\dot a_{i, {\mbox{\boldmath$\scriptstyle p$}} + \hbar {\mbox{\boldmath$\scriptstyle k$}}_1} & = & {i\over 2} \Omega_{1i} e^{- i \Delta_1 (t - t_0)} a_{1, {\mbox{\boldmath$\scriptstyle p$}}} + {i\over 2} \Omega_{2i} e^{- i \Delta_2 (t - t_0)} a_{2, {\mbox{\boldmath$\scriptstyle p$}} + \hbar ({\mbox{\boldmath$\scriptstyle k$}}_1 -  {\mbox{\boldmath$\scriptstyle k$}}_2)} \nonumber\\
& & - {i\over\hbar} e^{{i\over\hbar} {{p^{\prime 2}}\over{2m}} (t - t_0)} m \left(U - i \hbar g_a {\partial\over{\partial p^\prime_a}} - \hbar^2 U_{ab} {\partial\over{\partial p^\prime_a \partial p^\prime_b}} \right) \left(a_{i, {\mbox{\boldmath$\scriptstyle p$}^\prime}} e^{ - {i\over\hbar} {{p^{\prime 2}}\over{2m}} (t - t_0)}\right)_{{\mbox{\boldmath$\scriptstyle p$}}^\prime = {\mbox{\boldmath$\scriptstyle p$}} + \hbar {\mbox{\boldmath$\scriptstyle k$}}_1}  \label{DynInter} \\
\dot a_{2, {\mbox{\boldmath$\scriptstyle p$}} + \hbar ({\mbox{\boldmath$\scriptstyle k$}}_1 - {\mbox{\boldmath$\scriptstyle k$}}_2)} & = & {i\over 2}\Omega_{2i}^*  e^{i\Delta_2 (t - t_0)} a_{i, {\mbox{\boldmath$\scriptstyle p$}} + \hbar {\mbox{\boldmath$\scriptstyle k$}}_1} \nonumber\\
& & - {i\over\hbar} e^{{i\over\hbar} {{p^{\prime 2}}\over{2m}} (t - t_0)} m \left(U - i \hbar g_a {\partial\over{\partial p^\prime_a}} - \hbar^2 U_{ab} {\partial\over{\partial p^\prime_a \partial p^\prime_b}} \right) \left(a_{i, {\mbox{\boldmath$\scriptstyle p$}^\prime}} e^{ - {i\over\hbar} {{p^{\prime 2}}\over{2m}} (t - t_0)}\right)_{{\mbox{\boldmath$\scriptstyle p$}}^\prime = {\mbox{\boldmath$\scriptstyle p$}} + \hbar ({\mbox{\boldmath$\scriptstyle k$}}_1 - {\mbox{\boldmath$\scriptstyle k$}}_2)} \label{Dyna2}
\end{eqnarray}
where we defined the complex Rabi frequencies $\Omega_{1i} :=  {1\over \hbar}\langle i \mid {\mbox{\boldmath$d$}}\cdot{\mbox{\boldmath$E$}}\mid 1 \rangle e^{i\phi_1}$ and $\Omega_{2i} :=  {1\over \hbar}\langle i\mid {\mbox{\boldmath$d$}}\cdot{\mbox{\boldmath$E$}}\mid 2\rangle e^{i \phi_2}$ and 
\begin{eqnarray}
\Delta_1 & := & \omega_1^A + \omega_1 + {{{\mbox{\boldmath$p$}}^2}\over{2m\hbar}} - \omega_i^A - {{(\mbox{\boldmath$p$} + \hbar {\mbox{\boldmath$k$}}_1)^2}\over{2m\hbar}} \\
\Delta_2 & := & \omega_2^A + \omega_2 + {{(\mbox{\boldmath$p$} + \hbar ({\mbox{\boldmath$k$}}_1 - {\mbox{\boldmath$k$}}_2))^2}\over{2m\hbar}} - \omega_i^A - {{(\mbox{\boldmath$p$} + \hbar {\mbox{\boldmath$k$}}_1)^2}\over{2m\hbar}}
\end{eqnarray}
and where the dot means the time derivative and $()^*$ complex conjugation. 
Eqns (\ref{Dyna1}-\ref{Dyna2}) describe the dynamics of the set $S_{\mbox{\boldmath$\scriptstyle p$}} := \{\mid 1, {\mbox{\boldmath$p$}}\rangle, \mid i, \mbox{\boldmath$p$} + \hbar {\mbox{\boldmath$k$}}_1\rangle, \mid 2, \mbox{\boldmath$p$} + \hbar ({\mbox{\boldmath$k$}}_1 - {\mbox{\boldmath$k$}}_2)\rangle\}$ which is called a {\it closed set of coupled states}. 
Note that the dynamics, that is, the Schr\"odinger equation, maps $S_{\mbox{\boldmath$\scriptstyle p$}} \rightarrow S_{\mbox{\boldmath$\scriptstyle p$}}$ for all $\mbox{\boldmath$p$}$ if and only if the interaction is of polynomial form. 
A non-polynomial interaction leads to a non-local propagation in momentum space. 
This means that if one prepares a state with initial momentum $\mbox{\boldmath$p$}$ the above dynamics mixes only states which belong to $S_{\mbox{\boldmath$\scriptstyle p$}}$; no other states are involved. 
Each $S_{\mbox{\boldmath$\scriptstyle p$}}$ is a dynamically invariant subspace of the Hilbert space. 
Consequently, we have proved that only for polynomial interactions we still have a closed set of coupled states which, in the case of atom beam interferometry, will effectively reduce to a two-level system by adiabatic elimination of the intermediate state. 

We solve the dynamical equations (\ref{Dyna1}-\ref{Dyna2}) approximately by the following procedure: 
We assume that the time of the interaction between the laser and the atoms is small compared to the total flight time of the atoms inside the interferometer. 
In addition, during the laser-atom interaction the strength of the dipole interaction is much stronger than the gravitational interaction ($m U(\mbox{\boldmath$x$}) \ll \mbox{\boldmath$d$}\cdot\mbox{\boldmath$E$}$ for $\mbox{\boldmath$E$} \neq 0$). 
Thus it is possible to neglect the gravitational interaction for the laser-atom interaction zones. 
(It is possible to treat the influence of the acceleration on the beam splitting process \cite{LB95} which leads to small corrections in the phase shift of the order $10^{-2}$. 
Since we are more interested in principal questions we did not take this into account.)

\subsubsection{Beam splitters and mirrors}

For $U(\mbox{\boldmath$x$}) = 0$ we can adiabatically eliminate the intermediate level (see e.g. \cite{Moleretal92})
\begin{eqnarray}
i\hbar \dot a_{1, {\mbox{\boldmath$\scriptstyle p$}}} & = &  {1\over 2}\Omega_1^{\hbox{\scriptsize AC}} a_{1, {\mbox{\boldmath$\scriptstyle p$}}} + {1\over 2}\Omega_{\hbox{\scriptsize eff}} e^{i \delta (t - t_0)} a_{2, {\mbox{\boldmath$\scriptstyle p$}} + \hbar ({\mbox{\boldmath$\scriptstyle k$}}_1 -  {\mbox{\boldmath$\scriptstyle k$}}_2)} \label{2level1}\\
i\hbar \dot a_{2, {\mbox{\boldmath$\scriptstyle p$}} + \hbar ({\mbox{\boldmath$\scriptstyle k$}}_1 - {\mbox{\boldmath$\scriptstyle k$}}_2)} & = & {1\over 2}\Omega_{\hbox{\scriptsize eff}}^*  e^{- i \delta (t - t_0)} a_{1, {\mbox{\boldmath$\scriptstyle p$}}} +  {1\over 2} \Omega_2^{\hbox{\scriptsize AC}} a_{2, {\mbox{\boldmath$\scriptstyle p$}} + \hbar ({\mbox{\boldmath$\scriptstyle k$}}_1 - {\mbox{\boldmath$\scriptstyle k$}}_2)} \label{2level2}
\end{eqnarray}
where we introduced the definitions
\begin{eqnarray}
\Omega_1^{\hbox{\scriptsize AC}} & := & {{\Omega_{1i}\Omega_{1i}^*}\over{2 \Delta_1}}, \qquad \Omega_2^{\hbox{\scriptsize AC}} := {{\Omega_{2i}\Omega_{2i}^*}\over{2 \Delta_2}}, \quad \Omega_{\hbox{\scriptsize eff}} :=  {{\Omega_{1i}^*\Omega_{2i}}\over{2 \Delta_2}} \approx {{\Omega_{i2}\Omega_{i1}^*}\over{2 \Delta_1}} \\
\delta(\mbox{\boldmath$p$}) & := & \Delta_1 - \Delta_2 = \omega_1 - \omega_2 - \Biggl(\omega_{\hbox{\scriptsize hfs}} + {{\mbox{\boldmath$p$}}\over m}\cdot \mbox{\boldmath$k$} + {{\hbar {\mbox{\boldmath$k$}}^2}\over{2m}}\Biggr)\label{delta}
\end{eqnarray}
where $\omega_{\hbox{\scriptsize hfs}} := \omega_2^A - \omega_1^A$ is the frequency of the hyperfine transition and ${\mbox{\boldmath$k$}} := {\mbox{\boldmath$k$}}_1 - {\mbox{\boldmath$k$}}_2$. 
The ${\mbox{\boldmath$p$}}/m$-term describes the Doppler shift. 

Thus we are able to replace the three level system (\ref{Dyna1}-\ref{Dyna2}) effectively by the two level system (\ref{2level1},\ref{2level2}). 
The solutions of the above equations of motion (\ref{2level1},\ref{2level2}) is well known and can be used to define $\pi$ and $\pi/2$ pulses \cite{KasevichChu92} which serve as mirrors, beam splitters and recombiners. 
If the state is prepared at $t_0$ then we have for the $\pi$-pulse 
\begin{eqnarray}
a_{1, \mbox{\boldmath$\scriptstyle p$}}(t_+) & = & - i e^{i \left(\delta_0 + \delta(\mbox{\boldmath$\scriptstyle p$}) (t - t_0)\right)} a_{2, \mbox{\boldmath$\scriptstyle p$} + \hbar \mbox{\boldmath$\scriptstyle k$}}(t_-) \label{pi1}\\
a_{2, \mbox{\boldmath$\scriptstyle p$} + \hbar \mbox{\boldmath$\scriptstyle k$}}(t_+) & = & - i e^{- i \left(\delta_0 + \delta(\mbox{\boldmath$\scriptstyle p$}) (t - t_0)\right)} a_{1, \mbox{\boldmath$\scriptstyle p$}}(t_-)
\end{eqnarray}
and for the $\pi/2$-pulse 
\begin{eqnarray}
a_{1, \mbox{\boldmath$\scriptstyle p$}}(t_+) & = &  {1\over\sqrt{2}} \left(a_{1, \mbox{\boldmath$\scriptstyle p$}}(t_-) - i e^{i \left(\delta_0 + \delta(\mbox{\boldmath$\scriptstyle p$}) (t - t_0)\right)} a_{2, \mbox{\boldmath$\scriptstyle p$} + \hbar \mbox{\boldmath$\scriptstyle k$}}(t_-)\right) \\
a_{2, \mbox{\boldmath$\scriptstyle p$} + \hbar \mbox{\boldmath$\scriptstyle k$}}(t_+) & = & {1\over\sqrt{2}} \left(- i e^{- i \left(\delta_0 + \delta(\mbox{\boldmath$\scriptstyle p$}) (t - t_0)\right)} a_{1, \mbox{\boldmath$\scriptstyle p$}}(t_-) + a_{2, \mbox{\boldmath$\scriptstyle p$} + \hbar \mbox{\boldmath$\scriptstyle k$}}(t_-)\right)
\end{eqnarray}
Here $t$ is the time of the pulses which are assumed to be of infinitesimal short duration. 
$t_+$ and $t_-$ are the times immediately after and before the pulses ($t_\pm = t \pm \epsilon$ with $\epsilon\rightarrow 0$). 
$\delta_0$ is some additional constant phase depending on the laser. 

\subsubsection{The gravitational influence}

For the dark zones where no laser interacts with the atom only gravity is present. 
In these zones we have for our effective two level system
\begin{eqnarray}
\dot a_{r, {\mbox{\boldmath$\scriptstyle p$}}^{(r)}} & = & - {i\over\hbar} m \left[\left(U - {{\mbox{\boldmath$g$} \cdot {\mbox{\boldmath$p$}}^{(r)}}\over m} (t - t_0) + i \Delta U {{t - t_0}\over m} + U_{ab} {{\delta^{ac} \delta^{bd} p_c^{(r)} p_d^{(r)}}\over{m^2}} (t - t_0)^2\right) a_{r, {\mbox{\boldmath$\scriptstyle p$}}^{(r)}} \right.\nonumber\\
& & \left. + i \hbar \left(- g_a + U_{ab} {{\delta^{bd} p_d^{(r)}}\over m} (t - t_0) \right) {{\partial a_{r, {\mbox{\boldmath$\scriptstyle p$}}^{(r)}}}\over{\partial p_a}} - \hbar^2 U_{ab} {{\partial^2 a_{r, {\mbox{\boldmath$\scriptstyle p$}}^{(r)}}}\over{\partial p_a \partial p_b}} \right] \label{totdyn}
\end{eqnarray}
for $r = 1, 2$ and ${\mbox{\boldmath$p$}}^{(1)} = \mbox{\boldmath$p$}$ and ${\mbox{\boldmath$p$}}^{(2)} = \mbox{\boldmath$p$} + \hbar \mbox{\boldmath$k$}$. 
$\Delta$ is the Laplace operator and $\delta^{ab}$ the Kronecker symbol. 
The solution which is exact with respect to $U$ and $\mbox{\boldmath$g$}$ and is of first order in $U_{ab}$ is given by (see Appendix)
\begin{eqnarray}
a_{r, {\mbox{\boldmath$\scriptstyle p$}}^{(r)}}(t) & = & e^{-{i\over\hbar} \phi({\mbox{\boldmath$\scriptstyle p$}}^{(r)}, t, t_0)} \Biggl( A({\mbox{\boldmath$p$}}^{(r)}, t, t_0) a_{r, {\mbox{\boldmath$\scriptstyle p$}}^{(r)} - m \mbox{\boldmath$\scriptstyle g$} (t - t_0)}(t_0) \nonumber\\
& &  + B_a({\mbox{\boldmath$p$}}^{(r)}, t, t_0) {{\partial}\over{\partial p_a}} a_{r, {\mbox{\boldmath$\scriptstyle p$}}^{(r)} - m \mbox{\boldmath$\scriptstyle g$} (t - t_0)}(t_0) + C_{ab}(t, t_0) {{\partial^2}\over{\partial p_a \partial p_b}} a_{r, {\mbox{\boldmath$\scriptstyle p$}}^{(r)} - m \mbox{\boldmath$\scriptstyle g$} (t - t_0)}(t_0) \Biggr) \label{gravsol}
\end{eqnarray}
with
\begin{eqnarray}
\phi({\mbox{\boldmath$p$}}^{(r)}, t, t_0) & = & m U (t - t_0) - \mbox{\boldmath$g$}\cdot{\mbox{\boldmath$p$}}^{(r)} {1\over 2} (t - t_0)^2 + m {\mbox{\boldmath$g$}}^2 {1\over 6}(t - t_0)^3  \label{solphi} \\
A({\mbox{\boldmath$p$}}^{(r)}, t, t_0) & = & 1 + \Delta U  {1\over 2} (t - t_0)^2  - {i\over\hbar} U_{ab} \left({{\delta^{ac} \delta^{bd} p_c^{(r)} p_d^{(r)}}\over{m}} {1\over 3} (t - t_0)^3 - g^a \delta^{bd} p_d^{(r)} {1\over 8} (t - t_0)^4  \right.\nonumber\\
& & \left. \qquad\qquad + m g^a g^b {1\over{20}}(t - t_0)^5 \right) \nonumber\\ 
B_a({\mbox{\boldmath$p$}}^{(r)}, t, t_0) & = & U_{ab} \left( \delta^{bd}p_d^{(r)} {1\over 2} (t - t_0)^2 - m g^b {1\over 3} (t - t_0)^3 \right) \nonumber\\
C_{ab}(t, t_0) & = & i\hbar U_{ab} m (t - t_0) 
\end{eqnarray}

\subsubsection{Phase shift for the Kasevich-Chu interferometer}

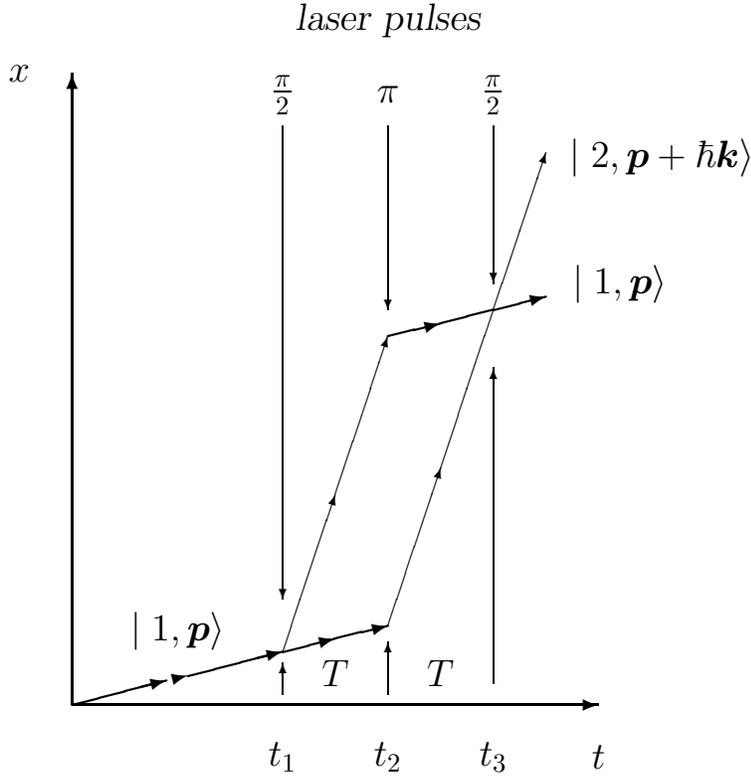
\begin{figure}
{\sf\Large
\unitlength1.4cm
\begin{picture}(8,8)
\put(3,1.5){\vector(1,3){0.5}}
\put(3.5,3){\vector(1,3){0.5}}
\put(4,1.75){\vector(1,3){0.5}}
\put(4.5,3.25){\vector(1,3){1}}
\put(6,0.5){\makebox(0,0){$t$}}
\put(3,0.5){\makebox(0,0){$t_1$}}
\put(4,0.5){\makebox(0,0){$t_2$}}
\put(5,0.5){\makebox(0,0){$t_3$}}
\put(3.5,1.3){\makebox(0,0){$T$}}
\put(4.5,1.3){\makebox(0,0){$T$}}
\put(0.5,7){\makebox(0,0){$x$}}
\put(3,1.1){\vector(0,1){0.3}}
\put(3,6.5){\vector(0,-1){4.5}}
\put(4,1.1){\vector(0,1){0.5}}
\put(4,6.5){\vector(0,-1){1.75}}
\put(5,1.2){\vector(0,1){3}}
\put(5,6.5){\vector(0,-1){1.5}}
\put(3,6.8){\makebox(0,0){${\pi\over 2}$}}
\put(4,6.8){\makebox(0,0){$\pi$}}
\put(5,6.8){\makebox(0,0){${\pi\over 2}$}}
\put(2,1.7){\makebox(0,0){$\mid 1, \mbox{\boldmath$p$} \rangle$}}
\put(6.2,5){\makebox(0,0){$\mid 1, \mbox{\boldmath$p$} \rangle$}}
\put(6.6,6.2){\makebox(0,0){$\mid 2, \mbox{\boldmath$p$} + \hbar \mbox{\boldmath$k$}\rangle$}}
\put(4,7.5){\makebox(0,0){\sf\sl laser pulses}}
\thicklines{\put(1,1){\vector(1,0){5}}
\put(1,1){\vector(0,1){6}}
\put(1,1){\vector(4,1){0.9}}
\put(1.9,1.225){\vector(4,1){0.2}}
\put(2.1,1.275){\vector(4,1){0.9}}
\put(3,1.5){\vector(4,1){0.5}}
\put(3.5,1.625){\vector(4,1){0.5}}
\put(4,4.5){\vector(4,1){0.5}}
\put(4.5,4.625){\vector(4,1){1}}
}
\end{picture}
}
\caption{Geometry of the atom beam interferometer.
Initially the atom beam is in the de-excited state $1$ with
momentum $\mbox{\boldmath$p$}$.
The first laser pulse splits the beam into two states 1 and 2 with
momentum $\mbox{\boldmath$p$}$ and $\mbox{\boldmath$p$} +
\hbar \mbox{\boldmath$k$}$. 
The second $\pi$-pulse acts as mirror and transforms state
1 into state 2 and vice versa.
The last $\pi/2$-pulse acts as recombiner.}
\label{fig1}
\end{figure}

Equations (\ref{pi1}-\ref{gravsol}) allow us to calculate the probability for the state $a_{2, {\mbox{\boldmath$\scriptstyle p$}} + \hbar {\mbox{\boldmath$\scriptstyle k$}}}(t_3)$ to leave the Kasevich-Chu interferometer (see Fig.\ \ref{fig1}). 
An initial state is prepared with $a_{2, {\mbox{\boldmath$\scriptstyle p$}} + \hbar {\mbox{\boldmath$\scriptstyle k$}}}(t_1) = 0$. 
At $t = t_1$ the first $\pi/2$-pulse acts as beam splitter and coherently splits this atom beam into states $a_{1, {\mbox{\boldmath$\scriptstyle p$}}}$ and $a_{2, {\mbox{\boldmath$\scriptstyle p$}} + \hbar {\mbox{\boldmath$\scriptstyle k$}}}$. 
During the time interval $T = t_2 - t_1$ both atomic states are subject to the gravitational interaction only where this interaction does not alter the internal states of the atom beam. 
At $t = t_2$ a $\pi$-pulse acts as mirror and transforms the state $a_{1, {\mbox{\boldmath$\scriptstyle p$}}}$ into the state $a_{2, {\mbox{\boldmath$\scriptstyle p$}} + \hbar {\mbox{\boldmath$\scriptstyle k$}}}$ and vice versa. 
Between $t_2$ and $t_3$ with $T = t_3 - t_2$ again only gravity acts and at $t = t_3$ a last $\pi/2$-pulse acts as a recombiner so that at $t_3$ both atom beams interfere with one another. 
The probability to detect $a_1$ or $a_2$ gives an interference pattern which is influenced by the gravitational interaction which is effective between the $\pi$ and $\pi/2$-pulses. 

The state $a_{2, {\mbox{\boldmath$\scriptstyle p$}} + \hbar {\mbox{\boldmath$\scriptstyle k$}}}(t_3)$ leaving the interferometer turns out to be
\begin{eqnarray}
a_{2, \mbox{\boldmath$\scriptstyle p$} + \hbar \mbox{\boldmath$\scriptstyle k$}}(t_3) & = & - {i\over 2} e^{i \delta^\prime} \left\{a_{1, \mbox{\boldmath$\scriptstyle p$} - 2 m \mbox{\boldmath$\scriptstyle g$} T}(t_1) + U_{ab} T^2 \left(p^b - {{3 \hbar k^b}\over 2} + {{13}\over 6} g^b m T\right) {{\partial a_{1, \mbox{\boldmath$\scriptstyle p$} - 2 m \mbox{\boldmath$\scriptstyle g$} T}(t_1)}\over{\partial p_b}} \right.\nonumber\\
& & \left. \qquad\qquad + 2 i \hbar  m T U_{ab}{{\partial^2 a_{1, \mbox{\boldmath$\scriptstyle p$} - 2 m \mbox{\boldmath$\scriptstyle g$} T}(t_1)}\over{\partial p_a \partial p_b}}\right.\nonumber\\
& & \left. - e^{i F} \left[a_{1, {\mbox{\boldmath$\scriptstyle p$}}}(t_1)  + U_{ab} T^2 \delta^{bc}\left(p_c + {{5 \hbar k_c}\over 2} + {{13}\over 6} g_c m T\right){{\partial a_{1, \mbox{\boldmath$\scriptstyle p$} - 2 m \mbox{\boldmath$\scriptstyle g$} T}(t_1)}\over{\partial p_b}} \right.\right.\nonumber\\
& & \left.\left. \qquad\qquad + 2 i  \hbar m T U_{ab} {{\partial^2 a_{1, \mbox{\boldmath$\scriptstyle p$} - 2 m \mbox{\boldmath$\scriptstyle g$} T}(t_1)}\over{\partial p_a \partial p_b}}\right]\right\}
\end{eqnarray}
with 
\begin{eqnarray}
F & = & - 2 T^2 k^a \left(g_a - {1\over 2} U_{ab} T \delta^{bc}\left( {1\over m} \left({1\over 2} \hbar k_c + p_c \right) - g_c {{37}\over{12}} T \right)\right)  - \delta_3 + 2 \delta_2 - \delta_1
\end{eqnarray}
We prepared the initial state $a_{2, {\mbox{\boldmath$\scriptstyle p$}} + \hbar {\mbox{\boldmath$\scriptstyle k$}}}(t_1) = 0$. 
$\delta^\prime$ is an overall phase which plays no role. 
The phases $\delta_1$, $\delta_2$, and $\delta_3$ come in via the $\pi$ and $\pi/2$-pulses. 
They can be chosen appropriately in the experiment. 
The fact that $a_1(t_1)$ depends on $\mbox{\boldmath$p$} - 2 m \mbox{\boldmath$g$} T$ while $a_2(t_3)$ depends on $\mbox{\boldmath$p$} + \hbar \mbox{\boldmath$k$}$ expresses the fact that between $t_1$ and $t_3$ the atoms are accelerated thus gaining an additional momentum $2 m \mbox{\boldmath$g$} T$. 

The observable quantity is the probability of finding the state $a_{2, {\mbox{\boldmath$\scriptstyle p$}} + \hbar {\mbox{\boldmath$\scriptstyle k$}}}(t_f)$ leaving the interferometer, that is $I_2 = \int |a_{2, \mbox{\boldmath$\scriptstyle p$} + \hbar \mbox{\boldmath$\scriptstyle k$}}(t_3)| d^3p$. 
We get 
\begin{equation}
I_2 = {1\over 2} \left[1 - \cos\phi\right] I_1 \end{equation}
with the total phase   
\begin{equation}
\phi := - k_a g^a T^2 - \delta_3 + 2 \delta_2 - \delta_1 + U_{ab} T^3 k^a\left(\left({{\hbar k^b}\over{2 m}} + {{\langle p^b \rangle_1}\over m} - g^b {{37}\over{12}} T \right) T + 4 \langle x^b \rangle_1 \right) \label{MZPhase}
\end{equation}
Here we used the expression for the incoming intensity and the mean values for the momentum and the position at the beam splitter, that is, at the moment of the first $\pi/2$-pulse. 
\begin{eqnarray}
I_1 & = & \int a^*_{1, \mbox{\boldmath$\scriptstyle p$}}(t_1) a_{1, \mbox{\boldmath$\scriptstyle p$}}(t_1) d^3p \nonumber\\ 
\langle p_a \rangle_1 & = & {1\over{I_1}} \int a^*_{1, \mbox{\boldmath$\scriptstyle p$}} p_a a_{1, \mbox{\boldmath$\scriptstyle p$}} d^3 p \\ 
\langle x^b \rangle_1 & = & {1\over{I_1}}\int a^*_{1, \mbox{\boldmath$\scriptstyle p$}}(t_1) i \hbar {{\partial a_{1, \mbox{\boldmath$\scriptstyle p$}}(t_1)}\over{\partial p_b}} d^3p \nonumber
\end{eqnarray}
For the discussion of (\ref{MZPhase}) we replace the momentum operator by the velocity operator $\widehat{\mbox{\boldmath$v$}} = \widehat{\mbox{\boldmath$p$}}/m$ which is possible according to Heisenberg's equation of motion. 
In addition, (\ref{MZPhase}) depends on the position expectation value $\langle \widehat{\mbox{\boldmath$x$}} \rangle_1$ of the quantum state of the atom beam at the time $t = t_1$. 
With regard to the expansion of the Newtonian potential it is most preferable to choose a coordinate system so that $\langle \widehat{\mbox{\boldmath $x$}} \rangle_1 = 0$. 
In this case 
\begin{equation}
\phi = \phi_{\hbox{\scriptsize grav}} - \delta_3 + 2 \delta_2 - \delta_1\quad\hbox{with}\quad \phi_{\hbox{\scriptsize grav}} = - 2 k^a T^2 \left(g_a - {1\over 2} U_{ab} T\left({{\hbar k^b}\over{2 m}} + \langle \widehat v^b \rangle_1 - g^b {{37}\over{12}} T\right)\right). \label{MZPhase1}
\end{equation}

Now we discuss this result:

\begin{enumerate}

\item  The result for the observable intensity uses only quantum notions like  wave vector, expectation value of the velocity and the time between the laser pulses. 
In no way is any path or enclosed area used. 
In fact, in the atom beam interferometric set up it is not possible to operationally define paths the atoms may take. 

\item The last point is also reflected by the fact that the result (\ref{MZPhase1}) is exact in a quantum sense: no classical approximation is used. 
The above result is also exact in the acceleration $g$ and is of first order in $U_{ab}$ which may be interpreted as the Newtonian part of the Riemannian curvature. 
The only approximations used are the approximation of the Newtonian potential to second order in the position and the approximation of the solution (\ref{gravsol}) in $U_{ab}$. 

\item The phase shift does not depend on the value $U({\mbox{\boldmath$x$}}_0)$ where ${\mbox{\boldmath$x$}}_0$ may be chosen as the position expectation value $\langle \widehat{\mbox{\boldmath$x$}}\rangle_1$ of the atom beam in the moment of the first $\pi/2$-pulse. 

\item The first term $- 2 k^a g_a T^2$ is the usual acceleration term which has been tested by Kasevich and Chu \cite{KasevichChu91}. 
This term is independent of the initial state of the atom. 
All momenta contribute to the effect in the same manner. 

Although our derivation and result are quantum mechanically exact we can rewrite the result {\it formally} in terms of classical notions: 
Introducing the "length" $\mbox{\boldmath$l$} = \langle \widehat{\mbox{\boldmath$v$}} \rangle_1 T$ and the "height" $\mbox{\boldmath$h$} = \hbar \mbox{\boldmath$k$} T/m$ the acceleration term reads $- m \mbox{\boldmath$h$} \cdot \mbox{\boldmath$g$} l/ \hbar |\langle\widehat{\mbox{\boldmath$v$}} \rangle_1|$. 
This result is identical to the famous COW phase shift (this has been observed also by Bord\'e \cite{Bordeprcomm95}). 
However, this is purely formal, because there is neither an operational separation of the atom beams \cite{Riehleetal91} nor an operational realisation of a "height" or a "length" of the interferometer. 
The only experimentally given entities (which is given in classical terms) are the mean velocity of the atom beam, the wave vector of the laser beams, and the time between the laser pulses. 

\item It is remarkable that in the case of uniform acceleration ($U_{ab} = 0$) the measured probability does not depend either on the mass or on the internal structure of the atom. 
This means that in this case the weak EP in the usual sense is exactly valid \cite{Borde92}. 
The mass dependence of the wave function (\ref{g-solution}) is no indication for a break-down of the weak EP ofr observable effects. 

However, in general the phase shift (\ref{MZPhase1}) depends on the second derivative of the Newtonian potential and so on the mean velocity $\langle \widehat{\mbox{\boldmath$v$}} \rangle_1$ of the atoms in the moment of the first $\pi/2$-pulse (For higher-order expansions of the Newtonian potential we get additional terms of the form $\langle \widehat v^a \widehat v^b \rangle_1$, etc.)  
$\langle \widehat{\mbox{\boldmath$v$}} \rangle_1$ is determined by the initial state $\psi(t_0)$ which is represented by $a_{r, {\mbox{\boldmath$\scriptstyle p$}}^{(r)}}(t_0)$. 
The occurrence of such terms makes it necessary to include the notion of the initial state into a formulation of a QEP. 
Note that the initial state does not contain any information about the mass parameter of the quantum system used for the interference; mass is determined only dynamically. 

Since the mass $m$ appears in the observable total phase shift (\ref{MZPhase1}) the weak EP is not valid. 
However, the corresponding term $U_{ab} T^3 \hbar k^a k^b/m$ is the only one  which is nonlinear in the transferred momentum $\hbar \mbox{\boldmath$k$}$. 
Terms linear in the transferred momentum are independent of the mass of the quantum system and therefore obey the weak EP. 
The non-linear term may be interpreted as follows: the phase shift measures the acceleration of the two split atomic waves. 
However, the second wave feels another acceleration than the first one. 
The acceleration the second wave feels depends on the `kick' $\hbar \mbox{\boldmath$k$}$ it gains. 
(In classical terms: the larger the `kick' $\hbar \mbox{\boldmath$k$}$, the larger the `separation' of the two waves). 
Therefore the position dependence of the acceleration, that is, a Newtonian curvature, leads to a mass dependence of interference experiments. 

However, and that is the main point in our discussion of the phase shift (\ref{MZPhase1}), the QEP is valid: 
We take as input the norm of the wave vector $k = (\mbox{\boldmath$k$} \cdot  \mbox{\boldmath$k$})^{1/2}$. 
Then the input independent phase shift is given by ($\bar{\mbox{\boldmath$k$}} = \mbox{\boldmath$k$}/k$)
\begin{equation}
\lim_{k\rightarrow 0} {1\over k} \phi_{\hbox{\scriptsize grav}} = - 2 \bar k^a T^2 \left(g_a + {1\over 2} U_{ab} T\left(\langle \widehat v^b \rangle_1 - g^b {{37}\over{12}} T\right)\right) \, .
\end{equation} 
This input independent phase shift is valid in an arbitrarily curved space-time. 
It depends on the initial state via the expectation value of the velocity at the first beam splitter and does not depend on the mass of the atoms. 
Therefore the {\it QEP is valid for atom beam interferometry} based on (\ref{Schroedinger}). 

The input independent phase shift is the phase shift for a small wave packet displacement normalised by a parameter which characterises the amount of displacement. (A more suggestive term might be small "wave packet separation". 
However, our result is an exact quantum result, so that the two atom states $a_1$ and $a_2$ in general cannot be separated in configuration space; in real experiments \cite{Riehleetal91} the beams are in fact not separated, but always overlap.) 

The most important consequence of the validity of the QEP is that also in interference experiments the influence of the Newtonian potential on the interference fringes (in the input independent form) does not depend on the used quantum matter at all. 
Therefore this influence is characterised completely by the environment, and not by the quantum system. 
Consequently, also in the quantum domain the experimental results show up the geometric nature of the gravitational interaction. 
This reflects the geometric nature of the minimal coupling procedure of the Klein-Gordon or of the Dirac equation which in the non-relativistic limit gives the Schr\"odinger equation we are dealing with. 
(The relativistic case which may give more direct insight into the relation between the coupling to gravity and QEP will be treated in a later work.) 

\item We have just seen that (\ref{Schroedinger}) automatically fulfills the QEP. 
In order to perform an experimental test of the QEP we have to introduce into (\ref{Schroedinger}) the inertial mass $m_i$ and (passive) gravitational mass $m_g$ where the quantity `mass' has to be interpreted quantum mechanically as some quantum property of the quantum state. 
The coerresponding shift (\ref{MZPhase1}) will be modified: 
\begin{equation}
\phi_{\hbox{\scriptsize grav}} = - 2 k^a T^2 \left({{m_g}\over{m_i}} g_a + {1\over 2} {{m_g}\over{m_i}} U_{ab} T\left({{\hbar k^b}\over{2 m_i}} + \langle \widehat v^b \rangle_1 - {{m_g}\over{m_i}} g^b {{37}\over{12}} T\right)\right). \label{MZPhase2}
\end{equation} 
The corresponding input independent phase shift then tests the equality of inertial and gravitational mass if one makes interference experiments with  different kinds of atoms. 
This means that the equality of inertial and gravitational mass indicates the validity of the QEP. 
Note that this can be tested even in curved space-time and that it is not necessary (and in general not possible) to transform away gravity. 

For a {\it homogeneous} gravitational field we get $\phi_{\hbox{\scriptsize grav}} = - 2 \mbox{\boldmath$k$} \cdot \mbox{\boldmath$g$} T^2 {{m_g}\over{m_i}}$ so that in this case the QEP is equivalent to the weak EP. 

A way to test the QEP is to take different types of atoms, say type $(a)$ and type $(b)$. 
Then one can test whether the phase shift for a given gravitational acceleration will be the same for all atoms. 
If this is done, then for the accuracy $\Delta\phi/\phi$ of the Kasevich-Chu interferometer of about $10^{-8}$ a null experiment will give the validity of the equivalence principle in the quantum domain in terms of the E\"otv\"os ratio $\eta = (m_g/m_i)^{(a)} - (m_g/m_i)^{(b)}$ to an accuracy of the same order $|\eta| \lesssim 10^{-8}$. 
Note that there are no approximation used. 
Such an experiment will be a {\it pure quantum test of the equivalence principle}. 
(For bulk matter the weak EP is verified to the order $10^{-12}$ \cite{Adelbergeretal94}). 
In general, the effect of $U_{ab}$ has to be taken into account by performing this experiment. 

There is another way to test the EP: For an accelerated observer we have to replace $m_g \mbox{\boldmath$g$}$ by $- m_i \mbox{\boldmath$a$}$ with the resulting phase shift $\phi_{\hbox{\scriptsize accel}} = 2 \mbox{\boldmath$k$} \cdot \mbox{\boldmath$a$} T^2$. 
Therefore, one can test the equality of inertial and gravitational mass first by carrying out an interference experiment in a homogeneous gravitational field $\mbox{\boldmath$g$}$ and second by doing the same experiment with an accelerating interferometer with $\mbox{\boldmath$a$} = - \mbox{\boldmath$g$}$. 
If the experimental results coincide then the accuracy of the interferometer gives an estimate for the ratio $m_g/m_i$. 

\item In the limit $\hbar\rightarrow 0$ the phase shift (\ref{MZPhase1}) does not depend on the mass $m$ so that in this case the QEP is valid. 
However, in our case this limit is not the classical limit of the Schr\"odinger equation describing the dynamics of the atomic quantum system. 
For $U_{ab} = 0$ there is no $\hbar$ in the phase shift at all and the corresponding phase shift is an exact quantum result without any approximation. 
$\hbar\rightarrow 0$ in our context means that the quantum nature of the momentum transfer is neglected. 

If we ignore the quantum nature of the momentum transfer by e.g. introducing formally the velocity $\delta \mbox{\boldmath$v$} = \hbar \mbox{\boldmath$k$}/m$ which the quantum system acquires, then there is no $\hbar$ at all in the phase shift. 
This means that if we describe the quantum system only by those quantities ($\mbox{\boldmath$k$}$, $T$, and $\langle \widehat{\mbox{\boldmath$v$}}\rangle_1$) which are given experimentally and if we disregard the quantum origin of $\delta \mbox{\boldmath$v$}$, then the equations for the phase shift do not contain any $\hbar$. 
$\hbar$ only enters equations describing quantum systems if one replaces experimentally given notions by classical ones (compare the derivation of a COW-like formula for atom beam interferometry above). 

\item The above QEP is violated by doing interferometry with e.g. a charged quantum system in an external electrostatic field. 
The respective interaction Hamiltonian reads $H = - e \phi = - e(\phi({\mbox{\boldmath$x$}}_0) + E_a({\mbox{\boldmath$x$}}_0) \delta x^a + {1\over 2} \partial_b E_a({\mbox{\boldmath$x$}}_0) \delta x^a \delta x^b + \ldots )$. 
Adding this interaction to the Hamiltonian (\ref{Schroedinger}) gives in first order an additional phase shift for the Kasevich-Chu interferometer  $\delta\phi_{\hbox{\scriptsize em}} = - 2 {e\over m}k^a T^2 \left(E_a + {1\over 2} \partial_b E_a T\left({{\hbar k^b}\over{2 m}} + \langle v^b \rangle_1 - E^b {{37}\over{12}} T\right)\right)$. 
The corresponding input independent phase shift clearly depends on the specific charge $e/m$ and therefore violates the QEP. 

\item Similar statements can be made and conclusions drawn for a Bord\'e interferometer, Ref.\ \cite{Borde89}. 

\end{enumerate}

\subsection{The QEP in Neutron interferometry}

\begin{figure}
{\epsfxsize=5in \epsfbox{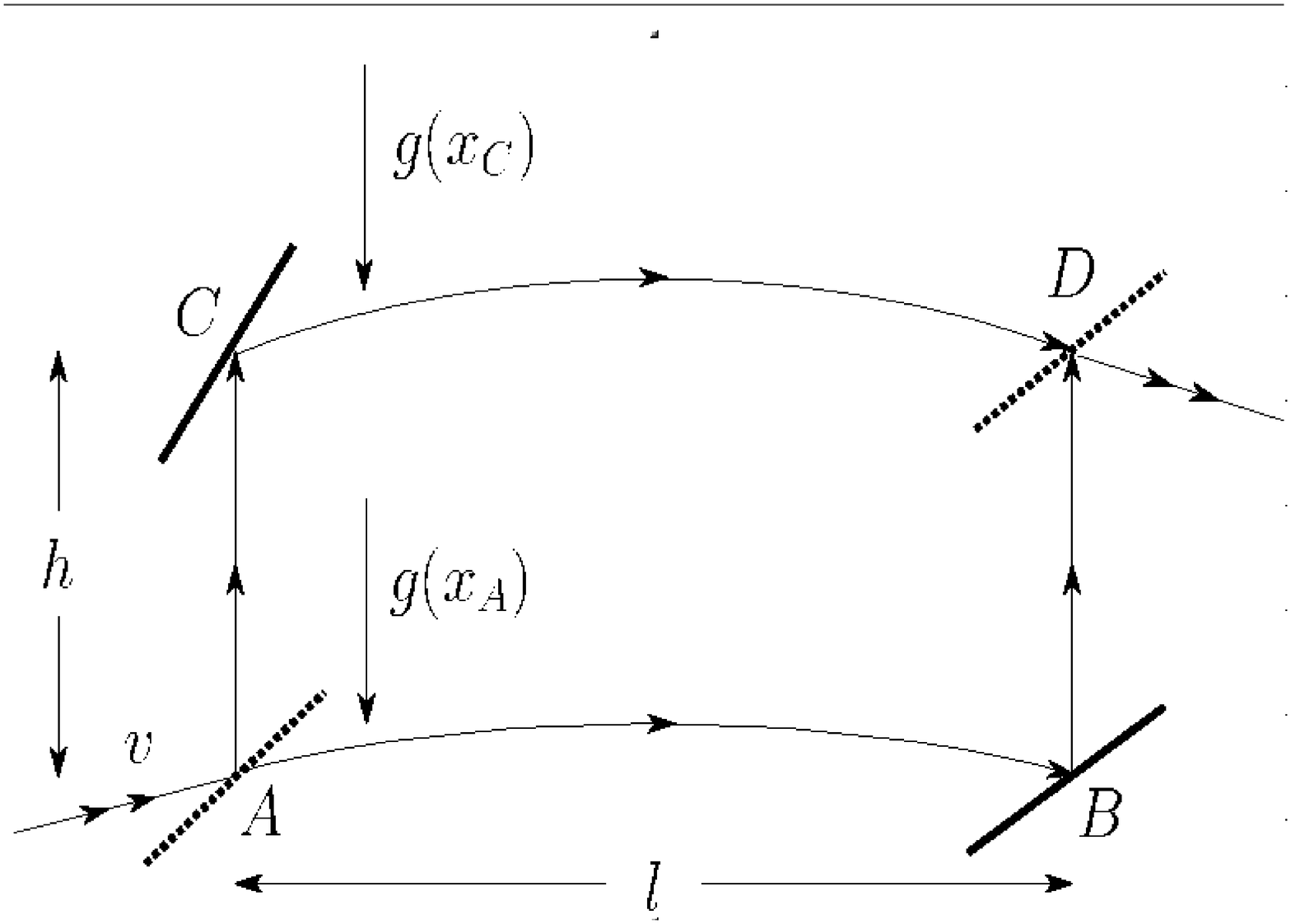}}
\caption{Mach-Zehnder geometry of a neutron interferometer
of length $l$ and
height $h$. 
The neutron beam hits the beam splitter $A$ with
velocity ${\mbox{\boldmath$v$}}$. 
$B$ and $C$ are mirrors and the beam is recombined at
$D$. 
Since the neutrons are in free fall between the beam splitter,
mirrors and recombiner, the path of the neutrons is curved.
The neutron beam feels a slightly different acceleration along
the upper and the lower path.}
\label{fig2}
\end{figure}

Neutron interferometry is most appropriately described by taking the Hamiltonian $H = {1\over{2 m}} \widehat{\mbox{\boldmath$p$}}^2 + m U(\mbox{\boldmath$x$})$ and performing a WKB approximation giving a Hamilton-Jacobi equation which can be solved for the modulus of the momentum $p(\mbox{\boldmath$x$}) = p \sqrt{1 + 2 m^2 U(\mbox{\boldmath$x$})/p^2}$ where $p$ is the momentum of the neutrons at the beam splitter. 
The phase shift is given by $\phi_{\hbox{\scriptsize grav}}^{\hbox{\scriptsize neutron}} = {1\over\hbar}\oint \mbox{\boldmath$p$}(\mbox{\boldmath$x$}) \cdot d \mbox{\boldmath$x$}$. 
We expand again the potential $U(\mbox{\boldmath$x$})$ to second order in the position. 
In the Mach-Zehnder geometry (see Fig.\ \ref{fig2}) we get for the intensity of one outgoing beam $I = I_0 (1 - \cos\phi_{\hbox{\scriptsize grav}}^{\hbox{\scriptsize neutron}})$ with the phase shift 
\begin{equation}
\phi_{\hbox{\scriptsize grav}}^{\hbox{\scriptsize neutron}} = {{m \mbox{\boldmath$g$}\cdot\mbox{\boldmath$h$} l}\over{\hbar v}} \left(1 + {{\mbox{\boldmath$g$}\cdot\mbox{\boldmath$h$}}\over{2 v^2}}\right) 
 + {{m U_{ab} h^a l}\over{2 \hbar v}}\left(h^b\left(1 + {{\mbox{\boldmath$g$}\cdot\mbox{\boldmath$h$}}\over{v^2}}\right) + {{g^b l^2}\over{6 v^2}}\right) \label{neutronPhase}
\end{equation}
where $l$ is the length, $\mbox{\boldmath$h$}$ the vector between the beam splitter $A$ and the mirror $C$, and $v = p/m$ the group velocity of the neutron beam at the beam splitter. 
We calculated (\ref{neutronPhase}) by taking into account the ballistic trajectory of the classical particle (curved lines in Fig.\ \ref{fig2}) where 
we oriented the distance $AB$ along constant potential $U(\mbox{\boldmath$x$})$ and the distances $AC$ and $BD$ parallel to the gravitational acceleration $\mbox{\boldmath$g$}({\mbox{\boldmath$x$}}_A) = \mbox{\boldmath$g$}({\mbox{\boldmath$x$}}_B)$. 
We calculated the trajectory between $A$ and $B$ using the acceleration $g({\mbox{\boldmath$x$}}_A)$ and the trajectory between $C$ and $D$ using $\mbox{\boldmath$g$}({\mbox{\boldmath$x$}}_C)$ with $g_a({\mbox{\boldmath$x$}}_C) = g_a({\mbox{\boldmath$x$}}_A) + U_{ab}({\mbox{\boldmath$x$}}_A) h^b$. 
The reflection at the mirrors $B$ and $C$ are adjusted so that both paths meet at the same point $D$ where the two beams recombine and interfere. 
This might physically be accomplished by bending the `ears' of the Bonse-Hart interferometer. 
(We neglected terms of the form $g^2 U_{ab}$.) 

It can be shown that the phase shift for neutron interferometry can also be cast into a form which exhibits the validity of our QEP. 
One way to show this is to treat the scattering of the neutrons at the crystal layers explicitely. 
In \cite{RauchPetraschek78} it was shown that the scattered neutron ray has the form $\psi = \alpha e^{i {\mbox{\boldmath$\scriptstyle \kappa$}} \cdot {\mbox{\boldmath$\scriptstyle x$}}} + \beta e^{i {\mbox{\boldmath$\scriptstyle \kappa$}}^\prime \cdot {\mbox{\boldmath$\scriptstyle x$}}}$ where ${\mbox{\boldmath$\kappa$}}$ is the wave vector of the incoming neutron wave and $\alpha$ and $\beta$ are some constants depending on the material of the crystal and on the wave vector. 
The wave vector ${\mbox{\boldmath$\kappa$}}^\prime$ of the scattered wave has to fulfill a Bragg condition which means that the direction of the outgoing wave vector is given by ${\mbox{\boldmath$\kappa$}}^\prime = {\mbox{\boldmath$\kappa$}} + {\mbox{\boldmath$G$}}$ where ${\mbox{\boldmath$G$}}$ is the reciprocal lattice vector near the Ewald sphere given by the momentum of the incoming neutron. 
We also have momentum conservation $\kappa = \kappa^\prime$. 
This wave vector corresponds to a group velocity ${\mbox{\boldmath$v$}}^\prime = \hbar {\mbox{\boldmath$\kappa$}}^\prime/m$. 
Therefore the reflected wave has a wave vector which is determined by the mirror used. 
This wave vector can be used to rewrite the phase shift formula for neutrons. 
First we replace $l$ by the interaction time $T$  which is the flight time of the horizontal part from $A$ to $B$ (see Fig.\ \ref{fig2}) within the neutron interferometer through $v = l/T$. 
In addition we can rewrite $\mbox{\boldmath$h$} = {\mbox{\boldmath$v$}}^\prime T^\prime = \hbar {\mbox{\boldmath$\kappa$}}^\prime T^\prime/m$ where $T^\prime$ is the flight time from $A$ to $C$ and from $B$ to $D$. 
$v^\prime$ is the velocity of the neutron after reflection at the mirror $A$ or $B$. 
According to the setup of the COW experiment ${\mbox{\boldmath$\kappa$}}$ lies in the direction of constant $U({\mbox{\boldmath$x$}})$ so that ${\mbox{\boldmath$\kappa$}} \cdot \mbox{\boldmath$g$} = 0$ and $\kappa^b U_{ab} = 0$. 
With ${\mbox{\boldmath$\kappa$}}^\prime = {\mbox{\boldmath$\kappa$}} + {\mbox{\boldmath$G$}}$ we have ${\mbox{\boldmath$\kappa$}}^\prime \cdot {\mbox{\boldmath$g$}} = {\mbox{\boldmath$G$}} \cdot {\mbox{\boldmath$g$}}$ and $\kappa^{\prime a} \kappa^{\prime b} U_{ab} = G^a G^b U_{ab}$. 
$\mbox{\boldmath$G$}$ plays the analogous role of $\mbox{\boldmath$k$}$ in atom beam interferometry. 
Therefore, in terms of wave mechanics only, that is, in terms of the interaction time $T$, the flight time $T^\prime$ and the reciprocal lattice vector ${\mbox{\boldmath$G$}}$ which is an experimentally given quantity depending on the preparation of the mirror, the phase shift is 
\begin{equation}
\phi_{\hbox{\scriptsize grav}}^{\hbox{\scriptsize neutron}} = \mbox{\boldmath$g$}\cdot\mbox{\boldmath$G$}\; T T^\prime \left(1 +  {{\hbar\;\mbox{\boldmath$g$}\cdot\mbox{\boldmath$G$}\; T^\prime}\over{2 m v^2}}\right) 
 + {1\over 2} U_{ab} G^a T T^\prime \left({{\hbar G^b T^\prime}\over m}\left(1 + {{\hbar\; \mbox{\boldmath$g$}\cdot\mbox{\boldmath$G$}\; T^\prime}\over{m v^2}}\right) + {{g^b T^2}\over{6}}\right) \label{neutronMZ}
\end{equation}

We add a few points of discussion:

\begin{enumerate}

\item This is only an approximate result. 
The approximations employed are the quasi--classical limit (approximation by plane waves), second order in the acceleration $\mbox{\boldmath$g$}$ and first order in $U_{ab}$. 

\item The first term is the COW phase shift \cite{COW75} which in our reformulation does not depend on $\hbar$ and on the mass $m$. 
The reason for this is that the above phase shift is a reformulation of the usual form of the phase shift in terms of notions of wave mechanics only. 
No notions from classical physics like path, enclosed area, height, or length of the interferometer are used. 
Only by introducing these classical notions do the mass and $\hbar$ enter. 
(Note that only the ratio $\hbar/m$ enters eqn (\ref{neutronPhase}) which is connected to the fact that only this ratio effectively appears in the Schr\"odinger equation (\ref{Schroedinger}). 
This is also of experimental significance \cite{WeissYoungChu94}.)  

\item The total phase shift contains terms which depend on the mass $m$ of the neutrons. 
However, the input independent result $\displaystyle \lim_{G \rightarrow 0} {1\over G}\phi_{\hbox{\scriptsize grav}}^{\hbox{\scriptsize neutron}} = T T^\prime  \bar{\mbox{\boldmath$G$}}\cdot \mbox{\boldmath$g$} + {1\over{12}} U_{ab} \bar G^a g^b T T^{\prime 3}$ does not depend on the mass of the neutrons anymore and therefore fulfills the QEP. 
(The limit $G \rightarrow 0$ can certainly not be realised by the Bonse-Hart single-crystal interferometer. 
However, using microfabricated gratings as neutron optical elements this limit can be reached.) 

\item In the case of uniform acceleration the phase shift depends on the mass in contrast to atom beam interferometry. 
This is due to the fact that the mirrors in atom beam interferometry do change the modulus of the momentum while the mirrors in neutron interferometry are momentum conserving. 

\item The phase shift (\ref{neutronMZ}) obeys the EP for $\hbar \rightarrow 0$. 
The corresponding remarks in the case of atom beam interferometry hold here too. 

\item It is clear that the phase shift (\ref{neutronPhase}) or (\ref{neutronMZ}) cannot be transformed away by choosing an accelerating reference frame. Consequently, the strong EP is not valid for neutron interferometry, neither. 

\end{enumerate}

Therefore we have proved that the QEP is valid for atom as well as for neutron interferometry. 
Next we show that it is also possible to use the QEP to determine the structure of the interaction of the quantum system with gravitational and inertial fields.

\section{Reconstruction of the interaction}

By means of the above result it is easy to reconstruct the gravitational interaction term in the Schr\"odinger equation. 
The procedure employed is exactly analogous to that used in Newtonian mechanics for structureless point particles:
First the force is defined as $\mbox{\boldmath$F$} = m \mbox{\boldmath$a$}$ where $m$ is the inertial mass and $\mbox{\boldmath$a$}$ the acceleration. 
The weak EP describes the experimental result that the path or the acceleration of the point particles does not depend on any parameters characterising the structureless point particle. 
Therefore the ratio $\mbox{\boldmath$F$}/m$ does not depend on the mass, charge etc. 
Consequently, the equation of motion of the point particle in a gravitational field is given by $\mbox{\boldmath$a$} = \mbox{\boldmath$f$} := \mbox{\boldmath$F$}/m$ where $\mbox{\boldmath$f$}$ does not depend on properties of the point particle. 
If we assume in addition that the force can be represented by a potential $\mbox{\boldmath$f$} = - \nabla U$ then we arrive at the equation of motion for point particles in a Newtonian potential: $\mbox{\boldmath$a$} = - \nabla U$. 

In a similar way we can proceed in the case of matter wave interferometry: 
We take as Hamiltonian 
\begin{equation}
H = {1\over{2m}}\widehat{\mbox{\boldmath$p$}}^2 + V(\mbox{\boldmath$x$}) + {1\over 2}(\mbox{\boldmath$V$}(\mbox{\boldmath$x$}) \cdot \widehat{\mbox{\boldmath$p$}} + \widehat{\mbox{\boldmath$p$}}\cdot \mbox{\boldmath$V$}(\mbox{\boldmath$x$})) + {1\over 2} W^{ab} \widehat p_a \widehat p_b \label{reconstructionHam}
\end{equation} 
where in addition to a general scalar interaction $V$ we have also inserted a vector interaction $\mbox{\boldmath$V$}$ and a tensor interaction $W^{ab}$; the latter, for simplicity, is assumed to be constant. 
To assure the hermiticity of the Hamiltonian we assume in additon that $V(\mbox{\boldmath$x$})$, $\mbox{\boldmath$V$}(\mbox{\boldmath$x$})$, and $W^{ab}$ are real. 
In addition, we have used a symmetrisation procedure. 
The mass $m$ is a constant characterising the type of quantum particles used (we do not consider spin). 
We expand the scalar interaction to second order in the position and the vector interaction to first order. 
Generalising the above procedure according to this generalised interaction we get as input invariant phase shift for an atom beam interferometric experiment 
\begin{eqnarray}
\lim_{k\rightarrow 0} {1\over k}\phi & = & 2 T^2 \bar k^a\left(- {1\over m} \partial_a V + {1\over{2m}}\partial_a\partial_b V T\left(\langle \widehat v^b \rangle_1 - {{37}\over{12 m}} T \delta^{bc}\partial_c V \right) \right. \nonumber\\
& & \left. + 2 \partial_a V^b \left(\delta_{bc}\langle\widehat v^c\rangle_1 + {1\over{4 m}} T \partial_b V \right) + \delta_{ac} W^{cb} \partial_b V\right)
\end{eqnarray}
An experimental verification of the QEP implies  that the above result is independent of the mass of the atom for all chosen values of $T$ and $\langle\widehat{\mbox{\boldmath$v$}}\rangle_1$. 
This forces one to conclude that $\partial_a V/m$, $\partial_a\partial_b V/m$, $\partial_a V^b$, and $m W^{ab}$ are independent of the mass (note that ${\mbox{\boldmath$V$}}({\mbox{\boldmath$x$}}_0)$ drops out of the result). 
Consequently, we can introduce $\partial_a U := {1\over m} \partial_a V$, $\partial_a\partial_b U = {1\over m} \partial_a\partial_b V$, and $U^{ab} := m W^{ab}$ where $U$ and $U^{ab}$ are independend of $m$. 
Thus, disregarding a constant, we have to insert $V = m U$ and $W^{ab} = U^{ab}/m$ as interaction term into the Schr\"odinger equation based on (\ref{reconstructionHam}): 
\begin{equation}
H = {1\over{2m}}(\delta^{ab} + U^{ab}) \widehat p_a \widehat p_b + m U(\mbox{\boldmath$x$}) + {1\over 2}(\mbox{\boldmath$V$}(\mbox{\boldmath$x$}) \cdot \widehat{\mbox{\boldmath$p$}} + \widehat{\mbox{\boldmath$p$}} \cdot \mbox{\boldmath$V$}(\mbox{\boldmath$x$}))  \label{HAnsatz}
\end{equation} 
where $U$, $\mbox{\boldmath$V$}$, and $U^{ab}$ do not depend on any properties of the quantum matter. 
Therefore we have derived that a universally coupling scalar potential must be multiplied by $m$, a universally coupling vector potential is not to be multiplied by $m$ while a universally coupling tensor potential must be divided by $m$. 
Thereby $m$ is exactly that parameter which appears as denominator in the kinetic part of the Schr\"odinger equation. 
The various interaction terms can be interpreted as follows: 

\begin{enumerate}

\item The scalar mass-independent potential $U$ is to be identified with the Newtonian potential. 
The special case $U(\mbox{\boldmath$x$}) = - \mbox{\boldmath$a$}\cdot \mbox{\boldmath$x$}$ can be interpreted as the transition of the Hamiltonian to an accelerated observer by means of the transformation $\mbox{\boldmath$x$} \rightarrow {\mbox{\boldmath$x$}}^\prime = \mbox{\boldmath$x$} - {1\over 2} \mbox{\boldmath$a$} t^2$ together with an additional unitary transformation. 

\item An interaction term of the form $V^a(\mbox{\boldmath$x$}) = \Lambda^a_b x^b$ with $\Lambda^a_b := \partial_b V^a({\mbox{\boldmath$x$}}_0)$ comes into the free Schr\"odinger equation by a transformation $x^a \rightarrow x^{\prime a} = \Lambda^a_b x^b t$. 
This is a general time-dependent affine transformation of the coordinates. 
In the case of an antisymmetric $\Lambda^a_b$ the contribution to the phase shift $\delta\phi = 2 T^2 k^a \Lambda^b_a \langle \widehat v_b\rangle_1$ is the usual Sagnac effect $\phi_{\hbox{\scriptsize Sagnac}} = 2 {m\over\hbar} \mbox{\boldmath$\omega$}\cdot\mbox{\boldmath$A$}$ rewritten in terms of wave mechanics only. 
Here we have to identify $\Lambda^a_b \leftrightarrow \mbox{\boldmath$\omega$}$ and $\mbox{\boldmath$A$}$ formally is the "area" enclosed by the "paths" of the atoms ($\mbox{\boldmath$A$} = \mbox{\boldmath$l$} \times \mbox{\boldmath$h$}$ with the "length" $\mbox{\boldmath$l$} = \langle\widehat{\mbox{\boldmath$v$}}\rangle_1 T$ and the "height" $\mbox{\boldmath$h$} = (\hbar \mbox{\boldmath$k$}/m) T$). 
The corresponding interaction term in (\ref{HAnsatz}) can be written as $- \mbox{\boldmath$\omega$} \cdot (\mbox{\boldmath$r$} \times \mbox{\boldmath$p$})$ and describes a special case of the coupling between rotation and total angular momentum (see \cite{Mashhoon88}). 
For $\mbox{\boldmath$k$}$ parallel to $\langle\widehat{\mbox{\boldmath$v$}}\rangle_1$ this effect vanishes. 
For a transformation where $\Lambda$ is a pure trace, $\Lambda^a_b = \Lambda \delta^a_b$ the transformation describes an expanding interferometer. 

Therefore the additional interaction ${1\over 2}(\mbox{\boldmath$V$}(\mbox{\boldmath$x$}) \cdot \widehat{\mbox{\boldmath$p$}} + \widehat{\mbox{\boldmath$p$}} \cdot \mbox{\boldmath$V$}(\mbox{\boldmath$x$})) = \Lambda^a_b {1\over 2} \{\widehat p_a, \widehat x^b\}$ describes the transition to an expanding, rotating or shearing frame of reference provided one demands that for the observed phase shifts the QEP is valid. 
Consequently, all input independent results due to changes of the frame of reference are independent of the properties of the atoms used. 

\item The tensor $\delta^{ab} + U^{ab}$ describes a modification of the Euclidean metric in the $t = const.$ hypersurface. 
However, by means of a coordinate transformation $x^a \rightarrow x^{\prime a} = K^a_b x^b$ with constant coefficients $K_a^b$ this tensor can be brought again to a $\delta^{ab}$-form. 

\end{enumerate}

Therefore we have derived by means of the QEP and of the phase shift the form of the interaction with gravitational and inertial fields in non-relativistic quantum mechanics. 

The same procedure applies also to neutron interferometry. 

Bearing in mind that the Schr\"odinger equation (\ref{Schroedinger}) can be obtained as the non-relativistic limit of the minimally coupled Klein-Gordon- or Dirac-equation, the above result may also be interpreted as an {\it operational} justification of the minimal coupling procedure in the non-relativistic limit. 
A more theoretical justification for the minimal coupling procedure has been given by Sciama \cite{Sciama}: 
On the level of a Lagrangian formulation of the dynamics of a matter field he  showed that invariance properties of the Lagrangian together with other conditions imply that the Lagrangian of the matter field should be coupled minimally to the underlying geometry.

\section{The QEP for expectation values}

For discussing the equivalence principle in the quantum domain it is also of interest to consider the equation of motion for the mean value of the position operator $\langle \widehat{\mbox{\boldmath$x$}} \rangle$. 
The bracket $\langle \widehat A \rangle = \langle \psi | \widehat A | \psi\rangle$ denotes the mean value of the observable $A$ for a quantum state $|\psi\rangle$. 
Because, for convenience, we are working in the Heisenberg picture, $|\psi\rangle$ is time-independent and can therefore be identified with the initially prepared state $|\psi_0\rangle$. 
We use the Schr\"odinger equation coupled to the Newtonian potential (\ref{Schroedinger}). 
For our discussion and derivation of the equation for the mean value of the position operator 
\begin{equation}
{{d^2}\over{dt^2}} \langle \widehat x^a \rangle = \delta^{ab} \langle {{\partial U(\widehat{\mbox{\boldmath$x$}}, t)}\over{\partial \widehat x^b}} \rangle \label{Schrquanf}
\end{equation}
we use the Heisenberg equation of motion
\begin{equation}
{{d^2}\over{dt^2}} \widehat x^a = \delta^{ab} {{\partial U(\widehat{\mbox{\boldmath$x$}}, t)}\over{\partial \widehat x^b}} \label{Schrquanf1}
\end{equation}
This equation (\ref{Schrquanf1}) is independent of any mass so that this relation cannot produce any mass dependence in (\ref{Schrquanf}). 
The only thing one has to do in going from (\ref{Schrquanf1}) to (\ref{Schrquanf}) is to multiply (\ref{Schrquanf1}) by the initial state $\langle\psi_0|$, resp. $|\psi_0\rangle$. 
However, the preparation of initial states is also independent of any parameter appearing in the dynamical equation. 
The initial state is just a function $\psi_0(\mbox{\boldmath$x$})$ which has to be in some Hilbert-space. 
Therefore we can prepare for atoms, neutron, electrons etc. the {\it same} initial state. 
Mass and charge are dynamical quantities. 

Therefore, the path of the mean value of the position operator depends only on the solution of the Heisenberg equation of motion (\ref{Schrquanf1}) with a given  potential function $U(\mbox{\boldmath$x$}, t)$ and on the choice on an initial state. 
Consequently, because neither the Heisenberg equation of motion for the position operator nor the initial state are mass dependent, {\it the QEP is valid for the equation for the mean value of the position operator}. 
For a given initial state the path $\mbox{\boldmath$x$}(t) = \langle \widehat{\mbox{\boldmath$x$}}(t) \rangle$ is independent of any mass parameter appearing in the Schr\"odinger equation. 
The mean values for the position operator of neutrons, atoms, and other neutral quantum systems follow the same path provided their initial states are the same. 
Interaction terms of the form $e \phi(\mbox{\boldmath$x$}, t)$ of course lead via the Heisenberg equation of motion to a dependence of the path $\mbox{\boldmath$x$}(t)$ on the specific charge $e/m$. 

In this connection we state once more that although in the case of a homogeneous gravitational field $U(\mbox{\boldmath$x$}) = \mbox{\boldmath$g$}(t) \cdot \mbox{\boldmath$x$}$ the solution of the Schr\"odinger equation in the Schr\"odinger picture depends on the mass, the path for the mean value of the position operator does not. 
The reason for this property is that the quantum state itself is not an observable quantity; the quantum state may depend on parameters which drop out of the expectation values (an overall constant phase is a special parameter which trivially drops out).  
Only expectation values (or mean values) are observables. 

Of course, the path $\mbox{\boldmath$x$}(t) = \langle \widehat{\mbox{\boldmath$x$}}(t) \rangle$ fulfilling (\ref{Schrquanf}) does not coincide with the respective path of a classical particle given by $\ddot{\mbox{\boldmath$x$}} = - \nabla U$, that is, $\langle \widehat{\mbox{\boldmath$x$}}(t) \rangle$ does not lead to an autonomous system. 
The difference of these two paths is of the order 
\begin{equation}
t^2\left(\partial_a U(\langle\widehat{\mbox{\boldmath$x$}}\rangle) - \langle {{\partial U(\widehat{\mbox{\boldmath$x$}})}\over{\partial \widehat x^b}} \rangle \right)  \label{diff}
\end{equation}
which vanishes in the classical limit. 
It is well known that for a potential $U(\mbox{\boldmath$x$}) = b_a x^a + b_{ab} x^a x^b$ (\ref{diff}) vanishes exactly so that the weak EP in its usual sense is valid. 

It can be shown in addition that the expectation value of each operator $\widehat A(\widehat x, \widehat v)$ using the solutions (\ref{gravsol}) does not depend on the mass $m$. 
This means that although the quantum state explicitely depends on the mass all expectation values of mass independent operators do not depend on the mass. 
One can also show that the time evolution of the uncertainties $\Delta x$ and $\Delta v$ do depend on the initial state only and not on the mass. 
Consequently, also for the uncertainties the QEP is valid. 

\section{Discussion}

The main aim of this paper was to find a useful notion of an equivalence principle in the quantum domain. 
We proposed a QEP which states that the output of a physical experiment in its input independent form only depends on the chosen initial state and not on the mass parameter appearing in the dynamical equation (in our case the Schr\"odinger equation) for the quantum system (atoms, neutrons, electrons, etc.). 
Consequently, even in the quantum domain we have found a way to extract results from the outcome of experiments where {\it gravity acts universally} in the sense that the gravitational influence does not depend on the mass of the quantum system. 
This means that even in the quantum domain the gravitational interaction is characterised completely by the environment, and not at all by the quantum system. 
The main implication of this is that even in the quantum domain the {\it gravitational interaction can be geometrised} in a well-defined manner. 

We can also interpret our approach in an operational manner. 
By means of our proposed QEP it is possible to decide from the outcome of e.g. an interference experiment whether there are interactions which act universally, i.e. which are independent of parameters characterising the used quantum system. 
These universal acting interaction fields may be called gravitational and inertial fields. 
Therefore our procedure is capable to {\it define gravitational fields in the quantum domain operationally}. 

It is of course desirable to extend the above considerations to the relativistic case. 
Also particles with spin should be included. 
It is presumed that also in these cases the above suggested formalism is useful. 

\section*{Acknowledgement}

It is a pleasure to thank Prof. J. Audretsch, Prof. Ch.J. Bord\'e and especially Prof. F.W. Hehl for discussions, Prof. R. Kerner for the hospitality in the LGCR at UPMC Paris, and the Deutsche Forschungsgemeinschaft for financial support. 

\section*{Appendix}

For $U_{ab} = 0$ equation (\ref{totdyn}) reduces to 
\begin{equation}
\dot a_{r, {\mbox{\boldmath$\scriptstyle p$}}^{(r)}}^0 = - {i\over\hbar} \left(m U - \mbox{\boldmath$g$} \cdot {\mbox{\boldmath$p$}}^{(r)} (t - t_0)\right) a_{r, {\mbox{\boldmath$\scriptstyle p$}}^{(r)}}^0 - m g_a {{\partial a_{r, {\mbox{\boldmath$\scriptstyle p$}}^{(r)}}^0}\over{\partial p_a}} 
\end{equation}
and can be solved exactly by
\begin{equation}
a_{r, {\mbox{\boldmath$\scriptstyle p$}}^{(r)}}^0(t) = e^{-{i\over\hbar} \phi({\mbox{\boldmath$\scriptstyle p$}}^{(r)}, t, t_0)}a_{r, {\mbox{\boldmath$\scriptstyle p$}}^{(r)} - m \mbox{\boldmath$\scriptstyle g$} (t - t_0)}(t_0) \label{g-solution}
\end{equation}
with $\phi({\mbox{\boldmath$p$}}^{(r)}, t, t_0)$ given by (\ref{solphi}).

To solve equation (\ref{totdyn}) approximately we treat $U_{ab}$ as small quantity. 
We make the ansatz 
\begin{equation}
a_{r, {\mbox{\boldmath$\scriptstyle p$}}^{(r)}}(t) = a^0_{r, {\mbox{\boldmath$\scriptstyle p$}}^{(r)}}(t) + a^1_{r, {\mbox{\boldmath$\scriptstyle p$}}^{(r)}}(t) \label{ansatz1}
\end{equation}
where $a^1_{r, {\mbox{\boldmath$\scriptstyle p$}}^{(r)}}(t)$ is small compared with $a^0(t)_{r, {\mbox{\boldmath$\scriptstyle p$}}^{(r)}}$ and insert this into (\ref{totdyn}), use the solution $a_0$ and neglect terms of the form $U_{ab} a^1_{r, {\mbox{\boldmath$\scriptstyle p$}}^{(r)}}(t)$. 
We finally get a partial differential equation for $a^1_{r, {\mbox{\boldmath$\scriptstyle p$}}^{(r)}}(t)$ with an inhomogeneity depending on $a^0_{r, {\mbox{\boldmath$\scriptstyle p$}}^{(r)} - m \mbox{\boldmath$\scriptstyle g$} (t - t_0)}(t_0)$. 
\begin{eqnarray}
\dot a^1_{r, {\mbox{\boldmath$\scriptstyle p$}}^{(r)}} & = & - {i\over\hbar} \left( m U - \mbox{\boldmath$g$} \cdot {\mbox{\boldmath$p$}}^{(r)} t\right)a^1_{r, {\mbox{\boldmath$\scriptstyle p$}}^{(r)}} 
- m g_a {{\partial a^1_{r, {\mbox{\boldmath$\scriptstyle p$}}^{(r)}}}\over{\partial p_a}} + \Delta U (t - t_0) e^{-{i\over\hbar} \phi({\mbox{\boldmath$\scriptstyle p$}}^{(r)}, t, t_0)}a_{r, {\mbox{\boldmath$\scriptstyle p$}}^{(r)} - m \mbox{\boldmath$\scriptstyle g$} (t - t_0)}(t_0) \nonumber\\
& & - {i\over\hbar} m U_{ab}  \left[ \left({{\delta^{ac} \delta^{bd} p_c^{(r)} p_d^{(r)}}\over{m^2}} (t - t_0)^2 \right.\right.\nonumber\\
& & \qquad\qquad \left.\left. - {{\delta^{bd} p_d^{(r)}}\over m} (t - t_0) g^a {1\over 2} (t - t_0)^2 + g^a g^b {1\over 4} (t - t_0)^4 \right)a_{r, {\mbox{\boldmath$\scriptstyle p$}}^{(r)} - m \mbox{\boldmath$\scriptstyle g$} (t - t_0)}(t_0) \right. \nonumber\\ 
& & \qquad \left. + i \hbar \left( {{\delta^{bd} p_d^{(r)}}\over m} (t - t_0) - g^a (t - t_0)^2 \right) {{\partial}\over{\partial p_a}} a_{r, {\mbox{\boldmath$\scriptstyle p$}}^{(r)} - m \mbox{\boldmath$\scriptstyle g$} (t - t_0)}(t_0) \right.\nonumber\\
& & \qquad \left. - \hbar^2  \left({{\partial^2}\over{\partial p_a \partial p_b}} a_{r, {\mbox{\boldmath$\scriptstyle p$}}^{(r)} - m \mbox{\boldmath$\scriptstyle g$} (t - t_0)}(t_0)\right) \right] e^{-{i\over\hbar} \phi({\mbox{\boldmath$\scriptstyle p$}}^{(r)}, t, t_0)} \label{pert1} 
\end{eqnarray}
Since the latter function has the phase $e^{-{i\over\hbar} \phi({\mbox{\boldmath$\scriptstyle p$}}^{(r)}, t, t_0)}$ we make for our unknown function the ansatz
\begin{equation}
a^1_{r, {\mbox{\boldmath$\scriptstyle p$}}^{(r)}}(t) = e^{-{i\over\hbar} \phi({\mbox{\boldmath$\scriptstyle p$}}^{(r)}, t, t_0)} b^1_{r, {\mbox{\boldmath$\scriptstyle p$}}^{(r)} - m \mbox{\boldmath$\scriptstyle g$} (t - t_0)}(t) \label{ansatz2}
\end{equation}
Inserting this into (\ref{pert1}) we get a differential equation for the perturbation $b^1_{r, {\mbox{\boldmath$\scriptstyle p$}}^{(r)} - m \mbox{\boldmath$\scriptstyle g$} (t - t_0)}(t)$:
\begin{eqnarray}
{{\partial^{\hbox{\scriptsize expl}}}\over{\partial t}} b^1_{r, {\mbox{\boldmath$\scriptstyle p$}}^{(r)} - m \mbox{\boldmath$\scriptstyle g$} (t - t_0)} & = & \Delta U (t - t_0) a_{r, {\mbox{\boldmath$\scriptstyle p$}}^{(r)} - m \mbox{\boldmath$\scriptstyle g$} (t - t_0)}(t_0) - {i\over\hbar} U_{ab}  \left[\left({{p^a p^b}\over{m}} (t - t_0)^2 - {{p^b}\over m} g^a {1\over 2} (t - t_0)^3 \right.\right.\nonumber\\
& & \left.\left. + m g^a g^b {1\over 4} (t - t_0)^4 \right)a_{1, {\mbox{\boldmath$\scriptstyle p$}} - m \mbox{\boldmath$\scriptstyle g$} (t - t_0)}(t_0) \right. \nonumber\\ 
& & \left. + i \hbar \left( p^b (t - t_0) - m g^b (t - t_0)^2 \right) {{\partial}\over{\partial p_a}} a_{r, {\mbox{\boldmath$\scriptstyle p$}}^{(r)} - m \mbox{\boldmath$\scriptstyle g$} (t - t_0)}(t_0) \right.\nonumber\\
& & \left. - m \hbar^2  {{\partial^2}\over{\partial p_a \partial p_b}} a_{r, {\mbox{\boldmath$\scriptstyle p$}}^{(r)} - m \mbox{\boldmath$\scriptstyle g$} (t - t_0)}(t_0) \right] 
\end{eqnarray}
This equation can be integrated whereby the integration is valid only for the explicit time-dependence, it does not act on the variable ${\mbox{\boldmath$p$}}^{(r)} - \mbox{\boldmath$g$} (t - t_0)$. 
We use in addition $a^1_{r, {\mbox{\boldmath$\scriptstyle p$}}^{(r)}}(t_0) = 0$ also $b^1_{r, {\mbox{\boldmath$\scriptstyle p$}}^{(r)}}(t_0) = 0$ and get  
\begin{eqnarray}
b^1_{r, {\mbox{\boldmath$\scriptstyle p$}}^{(r)} - m \mbox{\boldmath$\scriptstyle g$} (t - t_0)}(t) & = & {1\over 2} \Delta U (t - t_0)^2 a_{r, {\mbox{\boldmath$\scriptstyle p$}}^{(r)} - m \mbox{\boldmath$\scriptstyle g$} (t - t_0)}(t_0) - {i\over\hbar} U_{ab} \left[ \left({{\delta^{ac} \delta^{bd} p_c^{(r)} p_d^{(r)}}\over{m}} {1\over 3} (t - t_0)^3 - p^b g^a {1\over 8} (t - t_0)^4 \right.\right.\nonumber\\
& & \left.\left. + m g^a g^b {1\over{20}} (t - t_0)^5 \right)a_{r, {\mbox{\boldmath$\scriptstyle p$}}^{(r)} - m \mbox{\boldmath$\scriptstyle g$} (t - t_0)}(t_0) \right. \nonumber\\ 
& & \left. + i \hbar \left( \delta^{bc} p_c^{(r)} {1\over 2} (t - t_0)^2 - m g^b {1\over 3} (t - t_0)^3 \right) {{\partial}\over{\partial p_a}} a_{r, {\mbox{\boldmath$\scriptstyle p$}}^{(r)} - m \mbox{\boldmath$\scriptstyle g$} (t - t_0)}(t_0) \right.\nonumber\\
& & \left. - m \hbar^2 (t - t_0) {{\partial^2}\over{\partial p_a \partial p_b}} a_{r, {\mbox{\boldmath$\scriptstyle p$}}^{(r)} - m \mbox{\boldmath$\scriptstyle g$} (t - t_0)}(t_0) \right]  
\end{eqnarray}
This result together with (\ref{ansatz2}) and (\ref{ansatz1}) gives the solution (\ref{gravsol}).

\newpage

\end{document}